# Gd(III)-Gd(III) RIDME for In-Cell EPR Distance Determination

*Mykhailo Azarkh,§ Anna Bieber,§¶ Mian Qi,† Jörg W. A. Fischer,§ Maxim Yulikov,‡ Adelheid Godt,† Malte Drescher\*,§*

§Department of Chemistry, University of Konstanz, Universitätsstraße 10, 78457 Konstanz, Germany

¶Present address: Department of Molecular Structural Biology, Max Planck Institute of Biochemistry, Am Klopferspitz 18, 82152 Martinsried, Germany

†Faculty of Chemistry and Center for Molecular Materials (CM$_2$), Bielefeld University, Universitätsstraße 25, 33615 Bielefeld, Germany

‡Laboratory of Physical Chemistry, Department of Chemistry and Applied Biosciences, ETH Zurich, Vladimir-Prelog-Weg 2, 8093 Zurich, Switzerland

AUTHOR INFORMATION

**Corresponding Author**

\*E-mail: malte.drescher@uni-konstanz.de

In-cell distance determination by EPR reveals essential structural information about biomacromolecules under native conditions. We demonstrate that the pulsed EPR technique RIDME (relaxation induced dipolar modulation enhancement) can be utilized for such distance determination. The performance of in-cell RIDME has been assessed at Q band using stiff molecular rulers labelled with Gd(III)-PyMTA tags and microinjected into *X. laevis* oocytes. The overtone coefficients are determined to be the same for protonated aqueous solutions and inside cells. As compared to in-cell DEER (double electron-electron resonance, also abbreviated as PELDOR), in-cell RIDME features approximately 5 times larger modulation depth and does not show artificial broadening in the distance distributions due to the effect of pseudo-secular terms.

**TOC GRAPHICS**

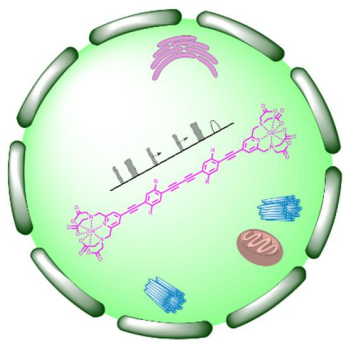

**KEYWORDS** RIDME, Xenopus laevis oocytes, in-cell distance determination, dipolar spectroscopy, Gd(III)-PyMTA, spin labels, in-cell EPR

Electron paramagnetic resonance (EPR) spectroscopy provides means to determine distances between magnetically coupled electron spins. By combination of pulsed EPR techniques with site-directed spin labelling, distances in the range up to 8 nm, and under the very special condition of a perdeuterated biomacromolecule up to 16 nm, can be determined.[1-10] Distance determination is based on the measurement of the dipolar coupling frequency, $\omega_{dd}$, which is inversely proportional to the cube of the distance between two magnetically coupled spins and can be performed in any kind of environment, including living cells. Because most cell components are diamagnetic, EPR-based distance determination inside cells is virtually background-free. Double electron-electron resonance (DEER or PELDOR)[11, 12] is the commonly used technique to perform such in-cell distance determination.[13-19] In-cell EPR distance determination put tight requirements on spin labels in terms of toxicity and stability. Complexes of Gd(III) with chelating ligands are ideal candidates for in-cell EPR and feature low toxicity, high stability, suitable spin relaxation times and no orientation selection.[18, 20-25] Despite all these favourable properties of Gd(III)-based spin labels, conventional Gd(III)-Gd(III) DEER with rectangular pulses is still not the ideal technique for in-cell distance determination because of its low modulation depth and artefacts at short distances, e.g. below 3.4 nm if Gd(III)-PyMTA is used as the spin label.[17, 18, 26, 27] Though it is possible to increase the modulation depth and to reduce the artefacts at short distances by using chirp pulses[28] and a dual-mode cavity,[29] these improvements can be achieved only with an additional comprehensive and expensive equipment of the EPR spectrometer with a high-power microwave amplifier, an arbitrary-waveform generator, a broad-band or a dual-mode cavity.

Alternatively, the downsides of conventional Gd(III)-Gd(III) DEER can be overcome in a related experiment abbreviated RIDME (relaxation induced dipolar modulation enhancement).[30-32]

RIDME is a single-frequency technique and makes use of relaxation-induced spin flips, whereas DEER is a two-frequency technique which detects at the observer frequency while flipping coupled spins in a controlled way by a $\pi$-pulse at the pump frequency. Thus RIDME is technically less demanding than DEER and has no limitations with respect to the excitation bandwidth. As an example, consider a rectangular pump pulse of 24 ns (excitation bandwidth of 40 MHz), which flips only a small part of the spins out of the 2-GHz-broad spectrum of Gd(III). While the contribution to the DEER signal is determined by this part of excited Gd(III) spins, the contribution to the RIDME signal is determined by the spins from the entire spectrum, which flip due to relaxation. Consequently, the modulation depth of in vitro Gd(III)-Gd(III) RIDME is increased by the factor of 10 compared to in vitro Gd(III)-Gd(III) DEER.[26, 33-35] These features of RIDME can also be beneficial for distance measurements inside cells. While the performance of Gd(III)-Gd(III) RIDME has recently been demonstrated for model systems in deuterated solvents at Q- and W-band frequencies,[33-35] the in-cell application remained elusive. To the best of our knowledge, the only in-cell RIDME experiment was reported for trityl spin labels, which are S=1/2 systems and have very different relaxation properties as compared to Gd(III) complexes.[36] Here, we utilize stiff molecular rulers spin labelled with a Gd(III)-PyMTA complex (Fig. 1A and B) and demonstrate that the 5-pulse dead-time free RIDME sequence[32] (Fig. 1C) is suitable for in-cell distance measurements between Gd(III)-based spin labels.

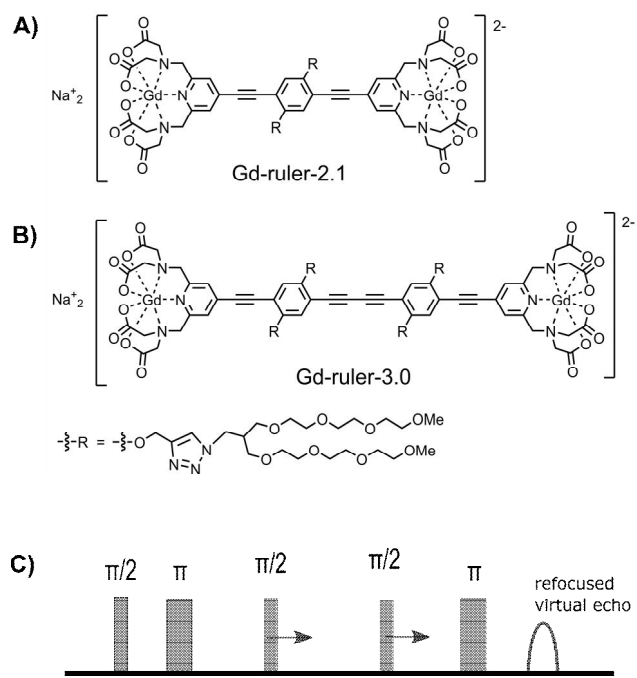

**Figure 1**. (A, B) Structural formulae of the Gd-ruler-2.1 and Gd-ruler-3.0, respectively. (C) Dead-time free RIDME pulse sequence.

In-cell RIDME implies that the measurement is performed in protonated media, which results in shorter relaxation times, as compared to deuterated solvents. Additionally, RIDME features a steeper signal decay than DEER because of the hyperfine spin diffusion. Therefore, accurate choice of experimental conditions with respect to relaxation and spin diffusion is crucial to realize in-cell RIDME. The measurement of relaxation times for Gd-ruler-3.0 in $H_2O$/glycerol (8/2 by volume) were performed in the temperature range between 10 and 30 K (Fig. S1). The phase memory time decreases from 1.4 to 0.8 μs and the ratio $T_2/T_1$ increases from 0.02 to 0.09 upon increasing the temperature. Though the highest ratio $T_2/T_1$ is favourable for RIDME, we performed all measurements at 10 K to have the longest phase-memory time. The pulse lengths were set at 8 and 16 ns for $\frac{\pi}{2}$ and $\pi$, respectively. The mixing time of 8 μs provided a reasonable tradeoff between the steepness of the RIDME background and the dipolar modulation depth. Figure 2 shows the results of in vitro RIDME measurements in frozen protonated aqueous solutions (8/2 mixture of $H_2O$ and glycerol) for two molecular rulers that bear Gd-PyMTA as the spin labels with distances of 2.1 and 3.0 nm between the two Gd(III) ions.[37-39] The RIDME form factors display modulation depths of 15% and 21% and visible oscillations, which indicate narrow distance distributions.

For determination of the interspin distances from the RIDME data, it was taken into account that relaxation-induced spin flips with $\Delta m \geq 1$ are possible for Gd(III) ions which have S = 7/2. Thus, apart from the fundamental frequency of the dipolar coupling $\omega_{dd}$, which corresponds to spontaneous spin flips with $\Delta m = 1$, overtones of this fundamental frequency occur, which originate from spin flips with $\Delta m > 1$. It has been shown for Gd(III) ions that only the first two overtones, i.e., $2\omega_{dd}$ and $3\omega_{dd}$, contribute significantly to the RIDME form factor and that the overtone fractions can be assumed constant for distances above 3 nm, but may vary if the distance is shorter.[35]

The RIDME data were processed with the OvertoneAnalysis software package, which explicitly includes the overtone fractions into the kernel function to extract the distance distribution by the Tikhonov regularization procedure.[35] The fractions, which are represented by the coefficients $P_1$, $P_2$, and $P_3$ (for the fundamental frequency $\omega_{dd}$ and its overtones at $2\omega_{dd}$ and $3\omega_{dd}$, respectively), were determined by their systematic variations towards a narrow single-peak distance distribution, as is expected for the studied Gd-rulers.[26, 35] We find the following overtone coefficients [$P_1$ $P_2$ $P_3$]: [0.68 0.21 0.11] for Gd-ruler-2.1 and [0.4 0.3 0.3] for Gd-ruler-3.0. The sets of coefficients for the Gd-rulers in the protonated solution are different from the ones determined in the deuterated solution (cf. [0.8 0.2 0.0] for Gd-ruler-2.1 and [0.51 0.40 0.09] for Gd-ruler-3.0).[35] From the comparison of the overtone coefficients for different solutions and different Gd-rulers, two observations follow. (i) Upon a decrease of the Gd(III)-Gd(III) distance but staying with the same solvent, $P_1$ increases while $P_2$ and $P_3$ decrease.[35] (ii) For the same Gd(III)-Gd(III) distance but changing from a protonated to a deuterated solvent, $P_1$ increases and $P_3$ decreases ($P_2$ stays about the same). The distance distributions for both molecular rulers in aqueous solutions (Fig. 2, right panel) were determined with the optimized overtone coefficients. The obtained distances are narrow and centered at 2.1 and 3.0 nm, in perfect agreement with the expectations.[35]

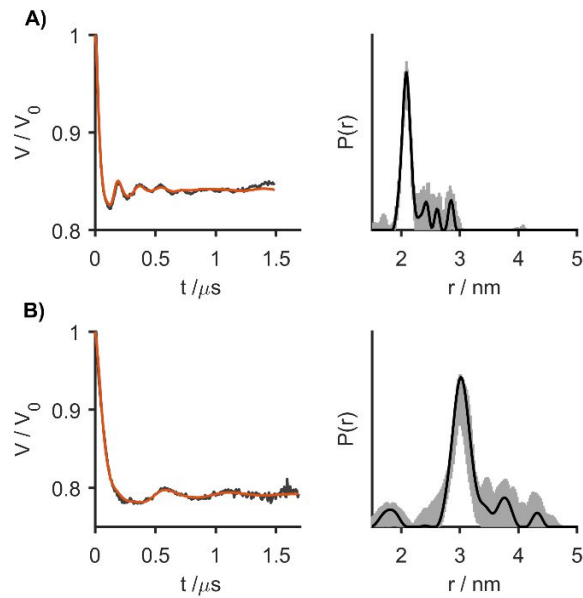

**Figure 2.** RIDME form factors (left) and corresponding distance distributions (right) obtained from frozen aqueous solutions of Gd-ruler-2.1 (A) and Gd-ruler-3.0 (B) in H$_2$O/glycerol (8/2

by volume). Gray areas in the right panels show the uncertainty range (as defined in SI) in the distance distributions.

After the proof-of-principle RIDME experiments in protonated media had been performed, Gd-ruler-3.0 was measured both in cell extract and inside oocytes of *X. laevis*. The redox stability of Gd(III)-based spin labels allows for long incubation times.[18,39] Thus, Gd-ruler-3.0 was incubated in the cell extract and inside oocytes for 2.5 h prior to shock freezing the sample for distance determination by RIDME. The form factors and the corresponding distance distributions are shown in Fig. 3. In both cases at least one oscillation is clearly visible in the form factor. The accurate measurement of the second and further weak oscillations in the in-cell RIDME data is difficult due to the lower signal-to-noise ratio. The modulation depth of the in-cell measurement (13 %, Fig. 3B) is smaller than in the aqueous solution (21 %, Fig. 2B) and in cell extract (17 %, Fig. 3A), which is most likely due to the endogenous Mn(II) present in the cells (approx. 10 µM).[18] Because endogenous Mn(II) ions contribute to the intensity of the RIDME signal but not to the modulation depth, the latter becomes diminished in in-cell measurements as compared with measurements in $H_2O$/glycerol and in cell extract, in full analogy with in-cell DEER.[18] The in-cell RIDME data for Gd-ruler-3.0 was analysed using the same overtone coefficients as found for the protonated aqueous solution (vide supra). The resulting distance distribution is in agreement with the distance distribution determined in aqueous solution. This suggests that the overtone coefficients determined with a protonated aqueous solution can be used to determine Gd(III)-Gd(III) distance distributions for both in-extract and in-cell RIDME measurements.

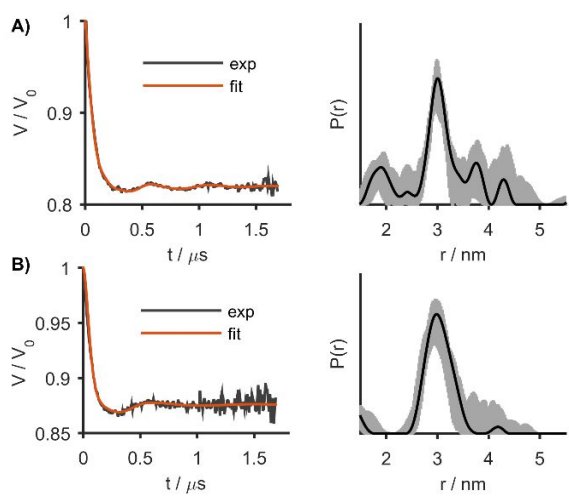

**Figure 3.** The RIDME form factors (left) and the corresponding distance distributions (right) for Gd-ruler-3.0 in cell extract (A) and in oocytes (B). Gray areas in the right panels show the uncertainty range (as defined in SI) in the distance distributions.

In general, in vitro RIDME measurements applied to Gd(III) spin labelled compounds provide larger modulation depths and more accurate distance distributions than in vitro DEER measurements.[34,35] In-cell measurements are performed under the challenging conditions of a protonated environment. In the presence of protons, the relaxation processes become faster and the hyperfine spin diffusion stronger. This results in a faster decay of the RIDME signal and shortens the length of the dipolar evolution time trace. If the experimental settings are chosen to maximize the length of the dipolar evolution time trace, i.e., low temperature and short mixing time, a RIDME time trace of at least 2 µs can be recorded, be it at the cost of the modulation depth. In this case, the modulation depths of in-cell and in vitro RIDME are in the range of 13-21 %, which is still less than the modulation depth of 50 % for RIDME measurements in deuterated solutions[33] but is a significant improvement over an in-cell DEER measurement using the same Gd-ruler-3.0 (a modulation depth of about 4 %).[18]

The overtone coefficients in a protonated aqueous solution and inside cells are the same for Gd-PyMTA spin labels at a given interspin distance. The distance distributions extracted from RIDME do not contain artefacts caused by pseudo-secular terms, as opposed to DEER.[20,26] We find that the overtone coefficients for aqueous solutions are significantly different for Gd(III)-Gd(III) distances of 2.1 and 3.0 nm, in agreement with what was reported for deuterated solvents.[35] This implies that for broad distance distributions, which span distances below and above 3 nm, the dependence of the overtone coefficients on the interspin distance has to be included in the processing of the RIDME data.

In conclusion, we demonstrated for the first time the performance of Gd(III)-Gd(III) RIDME for distance measurements inside cells. It is suggested that the overtone fractions remain the same for samples in aqueous solution and inside cells, being primarily determined by the presence of protons in the medium. At the current state, Gd(III)-Gd(III) RIDME can only be applied to distance distributions where the overtone fractions can be assumed constant. Further methodological developments are required to account for variable overtone fractions during data processing and also to slow down the decay of the RIDME signal during data acquisition. Along these lines, current developments of other Gd(III)-based spin labels[40-45] and the application of shaped pulses[46] are very promising. Giving its inherent advantages, i.e., precise distance determination and large modulation depth, RIDME has the potential to become a method of choice for in-cell distance measurements.

ASSOCIATED CONTENT

**Supporting Information** is available free of charge.

AUTHOR INFORMATION

**Notes**

The authors declare no competing financial interests.

ACKNOWLEDGMENT

This project has received funding from the European Research Council (ERC) under the European Union's Horizon 2020 research and innovation programme (Grant Agreement number: 772027 — SPICE — ERC-2017-COG). We thank Dr. Julia Cattani, Juliane Stehle, and Martina Adam for experimental contributions.

REFERENCES

(1) Berliner, L. J.; Eaton, S. S.; Eaton, G. R. (eds) *Biological Magnetic Resonance, Vol. 19. Distance Measurements in Biological Systems by EPR*; Kluwer Academics/Plenum Publishers: New York, 2000.

(2) Schiemann, O.; Prisner, T. F. Long-Range Distance Determination in Biomacromolecules by EPR Spectroscopy. *Q. Rev. Biophys.* **2007**, *102*, 377-390.

(3) Jeschke, G.; Polyhach, Y. Distance Measurements on Spin-Labelled Biomacromolecules by Pulsed Electron Paramagnetic Resonance. *Phys. Chem. Chem. Phys.* **2007**, *9*, 1895-1910.

(4) Jeschke, G. DEER Distance Measurements on Proteins, *Annu. Rev. Phys. Chem.* **2012**, *63*, 419-446.

(5) Schmidt, T.; Wälti, M. A.; Barber, J. L.; Husted, E. J.; Clore, G. M. Long Distance Measurements up to 160 A in the Groel Tetradecamer Using Q-Band DEER EPR Spectroscopy. *Angew. Chem. Int. Ed.* **2016**, *128* (51), 16137-16141.

(6) Altenbach, C.; Marti, T.; Khorana, H. G.; Hubbell, W. L. Transmembrane Protein Structure: Spin Labeling of Bachteriorhodopsin Mutants. *Science* **1990**, *248*, 1088-1092.

(7) Hubbell, W. L.; Cafiso, D. S.; Altenbach, C. Identifying Conformation Changes with Site-Directed Spin Labeling. *Nat. Struct. Biol.* **2000**, *7*, 735-739.

(8) Klare, J. P.; Steinhoff, H.-J. Spin Labeling EPR. *Photosynth. Res.* **2009**, *102*, 377-390.

(9) Fielding, A. J.; Concilio, M. G.; Heaven, G.; Hollas, M.A. New developments in spin labels for pulsed dipolar EPR. *Molecules.* **2014**, *19* (10), 16998-17025.

(10) Roser, P.; Schmidt, M. J.; Drescher, M.; Summerer, D. Site-Directed Spin Labeling of Proteins for Distance Measurements in Vitro and in Cells. *Org. Biomol. Chem.* **2016**, *14* (24), 5468-5476.

(11) Milov, A.; Ponomarev, A.; Tsvetkov, Y. D. Electron-Electron Double Resonance in Electron Spin Echo: Model Biradical Systems and the Sensitized Photolysis of Decalin. *Chem. Phys. Lett.* **1984**, *110* (1), 67-72.

(12) Pannier, M.; Veit, S.; Godt, A.; Jeschke, G.; Spiess, H. W. Dead-Time Free Measurement of Dipole-Dipole Interactions Between Electron Spins. *J. Magn. Reson.* **2000**, *142*, 331-340.

(13) Igarashi, R.; Sakai, T.; Hara, H.; Tenno, T.; Tanaka, T.; Tochio, H.; Shirakawa, M. Distance Determination in Proteins inside Xenopus laevis Oocytes by Double Electron-Electron Resonance Experiments. *J. Am. Chem. Soc.* **2010**, *132* (24), 8228-8229.

(14) Krstic, I.; Hänsel, R.; Romainczyk, O.; Engels, J. W.; Dötsch, V.; Prisner, T. F. Long-Range Distance Measurements on Nucleic Acids in Cells by Pulsed EPR Spectroscopy. *Angew. Chem. Int. Ed.* **2011**, *50*, 5070-5074.

(15)     Azarkh, M.; Okle, O.; Singh, V.; Seemann, I. T.; Hartig. J. S.; Dietrich, D. R.; Drescher, M. Long-Range Distance Determination in a DNA Model System inside Xenopus laevis Oocytes by In-Cell Spin-Label EPR. *ChemBioChem* **2011**, *12*, 1992-1995.

(16)     Azarkh, M.; Singh, V.; Okle, O.; Dietrich, D. R.; Hartig, J. S.; Drescher, M. Intracellular Conformations of Human Telomeric Quadruplexes Studied by Electron Paramagnetic Resonance Spectroscopy. *ChemPhysChem* **2012**, *13*, 1444-1447.

(17)     Martorana, A.; Bellapadrona, G.; Feintuch, A.; Di Gregorio, E.; Aime, S.; Goldfarb, D. Probing Protein Conformation in Cells by EPR Distance Measurements using $Gd^{3+}$ Spin Labeling. *J. Am. Chem. Soc.* **2014**, *136* (38), 13458-13465.

(18)     Qi, M.; Groß, A.; Jeschke, G.; Godt, A.; Drescher, M. Gd(III)-PyMTA Label Is Suitable for In-Cell EPR. *J. Am. Chem. Soc.* **2014**, *136*, 15366-15378.

(19)     Theillet, F.-X.; Binolfi, A.; Bekei, B.; Martorana, A.; Rose, H. M.; Stuiver, M.; Verzini, S.; Lorenz, D.; van Rossum, M,; Goldfarb, D.; Selenko, P. Structural Disorder of Monomeric alpha-Synuclein Persists in Mammalian Cells. *Nature* **2016**, *530*, 45-50.

(20)     Feintuch, A.; Otting, G.; Goldfarb, D. $Gd^{3+}$ Spin Labeling for Measuring Distances in Biomacromolecules: Why and How? *Methods in Enzymology* **2015**, *563*, 415-457.

(21)     Collauto, A.; Feintuch, A.; Qi, M.; Godt, A.; Meade, T.; Goldfarb, D. Gd(III) Complexes as Paramagnetic Tags: Evaluation of the Spin Delocalization over the Nuclei of the Ligand. *J. Magn. Reson.* **2016**, *263*, 156-163.

(22)     Weinmann, H. J.; Brasch, R. C.; Press, W. R.; Wesbey, G. E. Characteristics of Gadolinium-DTPA Complex – A Potential NMR Contrast Agent. *Am. J. Roentgenol.* **1984**, *142*, 619-624.

(23)     Bousquet, J. C.; Saini, S.; Stark, D. D.; Hahn, P. F.; Nigam, M.; Wittenberg, J.; Ferrucci, J. T. Gd-DOTA – Characterization of a New Paramagnetic Complex. *Radiology* **1988**, *166*, 693-698.

(24)     Cacheris, W. P.; Quay, S. C.; Rocklage, S. M. The Relationship between Thermodynamics and the Toxicity of Gadolinium Complexes. *M.R.I.* **1990**, *8*, 467-481.

(25)     Garbuio, L.; Zimmermann, K.; Häusslinger, D.; Yulikov, M. Gd(III) Complexes for Electron-Electron Dipolar Spectroscopy: Effects of Deuteration, pH and Zero Field Splitting. *J. Magn. Reson.* **2015**, *259*, 163-173.

(26)     Dalaloyan, A.; Qi, M.; Ruthstein, S.; Vega, S.; Godt, A.; Feintuch, A.; Goldfarb, D. Gd(III)-Gd(III) EPR Distance Measurements – The Range of Accessible Distances and the Impact of Zero Field Splitting. *PCCP* **2015**, *17*, 18464-18476.

(27)     Yang, Y.; Yang, F.; Li, X.-Y.; Su, X.-C.; Goldfarb, D. In-cell EPR Distance Measurements on Ubiquitin Labeled with a Rigid PyMTA-Gd(III) Tag. *J. Phys. Chem. B* **2019**, doi: 10.1021/acs.jpcb.8b11442

(28)     Doll, A.; Qi, M.; Wili, N.; Pribitzer, S.; Godt, A.; Jeschke, G. Gd(III)-Gd(III) Distance Measurements with Chirp Pump Pulses. *J. Magn. Reson.* **2015**, *259*, 153-162.

(29)     Cohen, M. R.; Frydman, V.; Milko, P.; Iron, M. A.; Abdelkader, E. H.; Lee, M. D.; Swarbrick, J. D.; Raitsimring, A.; Otting, G.; Graham, B.; Feintuch, A.; Goldfarb, D. Overcoming Artificial Broadening in $Gd^{3+}$-$Gd^{3+}$ Distance Distributions Arising from Dipolar Pseudo-Secular Terms in DEER Experiments. *PCCP* **2016**, *18*, 12847-12859.

(30)     Kulik, L. V.; Dzuba, S. A.; Grigoryev, I. A.; Tsvetkov, Yu, D. Electron Dipole-Dipole Interaction in ESEEM of Nitroxide Biradicals. *Chem. Phys. Lett.* **2001**, *343*, 315-324.

(31)     Astashkin, A. V. Mapping the Structure of Metalloproteins with RIDME. *Methods in Enzymology* **2015**, *563*, 251-284.

(32)     Milikisyants, S.; Scarpelli, F.; Finiguerra, M. G.; Ubbink, M.; Huber, M. A Pulsed EPR Method to Determine Distances between Paramagnetic Centers with Strong Spectral Anisotropy and Radicals: The Dead-Time Free RIDME Sequence. *J. Magn. Reson.* **2009**, *201*, 48-56.

(33)	Razzaghi, S.; Qi, M.; Nalepa, A. I.; Godt, A.; Jeschke, G.; Savitsky, A.; Yulikov, M. RIDME Spectroscopy with Gd(III) Centers. *J. Phys. Chem. Lett.* **2014**, *5*, 3970-3975.

(34)	Collauto, A.; Frydman, V.; Lee, M. D.; Abdelkader, E. H.; Feintuch, A.; Swarbrick, J. D.; Graham, B.; Otting, G.; Goldfarb, D. RIDME Distance Measurements Using Gd(III) Tags with a Narrow Central Transition. *Phys. Chem. Chem. Phys.* **2016**, *18*, 19037-19049.

(35)	Keller, K.; Mertens, V.; Qi, M.; Nalepa, A. I.; Godt, A.; Savitsky, A.; Jeschke, G.; Yulikov, M. Computing Distance Distributions from Dipolar Evolution Data with Overtones: RIDME Spectroscopy with Gd(III)-based Spin Labels. *Phys. Chem. Chem. Phys.* **2017**, *19*, 17856-17876.

(36)	Jassoy, J. J.; Berndhäuser, A.; Duthie, F.; Kühn, S. P.; Hagelueken, G.; Schiemann, O. Versatile Trityl Spin Labels for Nanometer Distance Measurements on Biomolecules In Vitro and within Cells. *Angew. Chem. Int. Ed.* **2017**, *56*, 177-181.

(37)	Qi, M.; Hülsmann, M.; Godt, A. Spacers for Geometrically Well-Defined Water-Soluble Molecular Rulers and Their Application. *J. Org. Chem.* **2016**, *81(6)*, 2549-2571.

(38)	Qi, M.; Hülsmann, M.; Godt, A. Synthesis and Hydrolysis of 4-Chloro-PyMTA and 4-Iodo-PyMTA Esters and Their Oxidative Degradation with Cu(I/II) and Oxygen. *Synthesis* **2016**, *48*, 3773-3784.

(39)	Yang, Y.; Yang, F.; Li, X.-Y.; Su, X.-C.; Goldfarb, D. In-Cell EPR Distance Measurements on Ubiquitin Labeled with a Rigid PyMTA-Gd(III) Tag. *J. Phys. Chem. B* **2019**, doi: 10.1021/acs.jpcb.8b11442.

(40)	Clayton, J. A.; Qi, M.; Godt, A.; Goldfarb, D.; Han, S.; Sherwin, M. S. $Gd^{3+}$-$Gd^{3+}$ Distances Exceeding 3 nm Determined by Very High Frequency Continuous Wave Electron Paramagnetic Resonance. *Phys. Chem. Chem. Phys.* **2017**, *19*, 5127-5136.

(41)	Mascali, F. C.; Ching, H. Y. V.; Rasia, R. M.; Un, S.; Tabares, L. Using Genetically Encodable Self-Assembling $Gd^{III}$ Spin Labels to Make In-Cell Nanometric Distance Measurements. *Angew. Chem. Int. Ed.* **2016**, *55* (37), 11041-11043.

(42)     Yang, Y.; Yang, F.; Gong, Y.-J.; Chen, J.-L.; Goldfarb, D.; Su, X.-C. A Reactive, Rigid Gd$^{III}$ Labeling Tag for In-Cell EPR Distance Measurements in Proteins. *Angew. Chem. Int. Ed*. **2017**, *56*, 2914-2918.

(43)     Welegedara, A. P.; Yang, Y.; Lee, M. D.; Swarbrick, J. D.; Huber, T.; Graham, B.; Goldfarb, D.; Otting, G. Double-Arm Lanthanide Tags Deliver Narrow Gd$^{3+}$-Gd$^{3+}$ Distance Distributions in DEER Measurements. *Chem. Eur. J*. **2017**, *23* (48), 11694-11702.

(44)     Mahawaththa, M. C.; Lee, M. D.; Giannoulis, A.; Adams, L. A.; Feintuch, A.; Swarbrick, J. D.; Graham, B.; Nitsche, C.; Goldfarb, D.; Otting, G. Small Neutral Gd(III) Tags for Distance Measurements in Proteins by Double Electron-Electron Resonance Experiments. *Phys. Chem. Chem. Phys*. **2018**, *20*, 23535-23545.

(45)     Yang, Y.; Yang, F.; Gong, Y.-J.; Bahrenberg, T.; Feintuch, A.; Su, X.-C.; Goldfarb, D. High Sensitivity In-Cell EPR Distance Measurements on Proteins Using an Optimized Gd(III) Spin Label. *J. Phys. Chem. Lett*. **2018**, *9*, 6119-6123.

(46)     Doll, A.; Qi, M.; Pribitzer, S.; Wili, N.; Yulikov, M.; Godt, A.; Jeschke, G. Sensitivity Enhancement by Population Transfer in Gd(III) Spin Labels. *Phys. Chem. Chem. Phys*. **2015**, *17*, 7334-7344.

# Gd(III)-Gd(III) RIDME for EPR In-Cell Distance Determination

*Mykhailo Azarkh,§ Anna Bieber,§¶ Mian Qi,† Jörg W. A. Fischer,§ Maxim Yulikov,‡ Adelheid Godt,† Malte Drescher\*,§*

§Department of Chemistry and Konstanz Research School Chemical Biology, University of Konstanz, Universitätsstraße 10, 78457 Konstanz, Germany

¶Present address: Department of Molecular Structural Biology, Max Planck Institute of Biochemistry, Am Klopferspitz 18, 82152 Martinsried, Germany

†Faculty of Chemistry and Center for Molecular Materials (CM$_2$), Bielefeld University, Universitätsstraße 25, 33615 Bielefeld, Germany

‡Laboratory of Physical Chemistry, Department of Chemistry and Applied Biosciences, ETH Zurich, Vladimir-Prelog-Weg 2, 8093 Zurich, Switzerland

Corresponding Author

\*E-mail: malte.drescher@uni-konstanz.de

## Supporting Information

**Experimental details**

Sample preparation.

The syntheses of Gd-ruler-2.1 and Gd-ruler-3.0 have been reported elsewhere.[1,2] Oocytes from *X. laevis* (stage V/VI) were purchased from EcoCyte Bioscience, Caustrop/Rauxel, Germany. The cell extract from the oocytes was prepared as described previously.[3]

All samples were prepared from aqueous solutions of Gd-rulers. Briefly, a 5 mM solution of Gd-ruler-2.1 in $D_2O$ (pH ~ 8.0, containing ca. 37 mM NaCl) or a 5 mM solution of Gd-ruler-3.0 in $D_2O$ (pH ~ 7.0, containing ca. 0.5 mM sodium trifluorocetat and ca. 30 mM NaCl) was lyophilized. The residual powder was dissolved in a predefined amount of $H_2O$ (Milli Q).

For in-vitro samples, a 200 μM solution of the Gd-rulers was prepared in a mixture of $H_2O$ (Milli Q) and glycerol (8:2, v/v).

For the in-extract sample, 1 μl of a 5 mM stock solution of Gd-ruler-3.0 in $H_2O$ was mixed with 24 μl of cell extract to produce a final concentration of 200 μM. No glycerol was added. The samples were transferred into quartz capillaries 1.6/1.0 mm o.d./i.d. (Bruker Biospin), shock-frozen in liquid nitrogen and stored at -80 °C until further use.

For the in-cell sample, 5 mM stock solution of Gd-ruler-3.0 in $H_2O$ was microinjected into oocytes (50 nl per oocyte) using a Nanojet II automatic nanoliter injector with fitting micromanipulator MM33 (DRUMMOND) to give a final in-cell concentration of 200 μM. After 2.5 h incubation at room temperature, 30 oocytes loaded with the Gd-ruler were transferred into a quartz tube 3.0/2.0 mm o.d./i.d., shock-frozen in liquid nitrogen and stored at -80 °C until further use.

EPR spectroscopy

All EPR measurements were performed at Q band on an Elexsys E580 spectrometer equipped with an arbitrary waveform generator (Bruker Biospin) and a 150 W TWT amplifier (Applied Systems Engineering). Temperature control was realized with the cryogen-free system consisting of a helium compressor F-70H (Sumitomo Cryogenics of America), a cryocooler ColdEdge (CE-FLEX-4K-0100, Bruker), and a MercuryITC (Oxford Instruments). A Q-band probehead with access for 3 mm tubes (ER5106QT-2, Bruker) was used for both 1.6 mm and 3 mm (o.d.) sample tubes. All the measurements were performed in the dip of the overcoupled resonator at 34 GHz. Rectangular pulses were used with the lengths fixed at 8 and 16 ns for $\pi/2$ and $\pi$, respectively.

The primary echo decay ($\pi/2 - \pi - echo$) was recorded by increasing the distance between the two microwave pulses with an 8 ns step. The length of the primary echo decay was 15 μs. The inversion recovery ($\pi - \pi/2 - \pi - echo$) was recorded by increasing the distance between the first and the second pulse with a 100 ns step. The length of the inversion recovery was 200 ms, In both experiments, the magnetic field was set to 12207 G, which corresponded to the maximum of the absorption curve.

In the RIDME measurement, a 5-pulse scheme was used and an 8-step phase cycle was applied during the acquisition of the refocused virtual echo.[4] The magnetic field was set to the maximum of the absorption spectrum of Gd-PyMTA label. A time delay between the first and the second pulse was 300 ns, a mixing time was 8 μs. The total length of the RIDME dipolar evolution was 2 μs, which was acquired with an 8 ns step.

The RIDME time traces were processed with the OvertoneAnalysis software package for Matlab, which includes a modified kernel function to account for overtone harmonics.[5] The background contributions to every RIDME time trace were approximated by a stretched exponential and eliminated by division. The distance distribution was extracted from the resulting form factor by Tikhonov regularization, where the fractions of overtones were fixed at $P_1 = 0.4$, $P_2 = 0.3$, and $P_3 = 0.3$ (for Gd-ruler-3.0) or at $P_1 = 0.69$, $P_2 = 0.21$, and $P_3 = 0.10$ (for Gd-ruler-2.1). Subsequently, the distance distribution at the optimum alpha-value (ranging from 7 to 39) was subjected to the validation. The validation by varying the background start, the modulation depth, and adding white noise at the level 1.2 to give the uncertainty regions for the distance distribution, which were plotted as grey area on top of the distance distribution.[5]

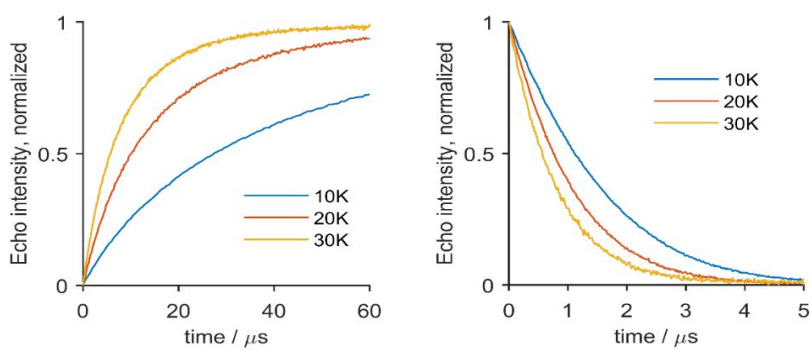

**Figure S1**. Inversion recovery (left) and primary echo decay (right) for Gd-ruler-3.0 in aqueous solution at different temperatures.

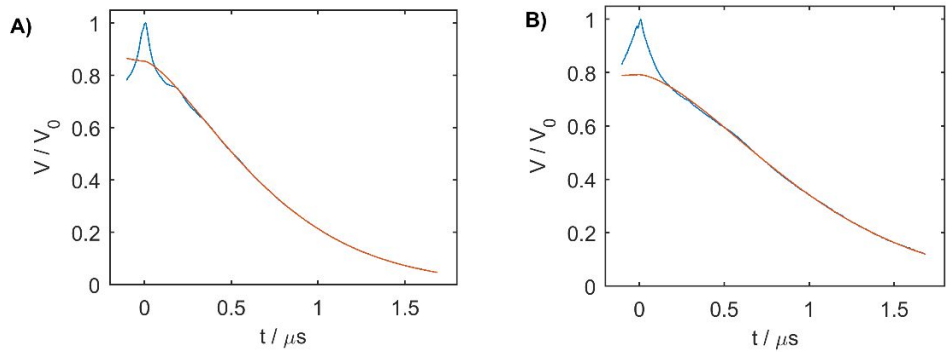

**Figure S2**. RIDME time traces (blue) and the background function (red) for Gd-ruler-2.1 (A) and Gd-ruler-3.0 (B) in frozen aqueous solutions.

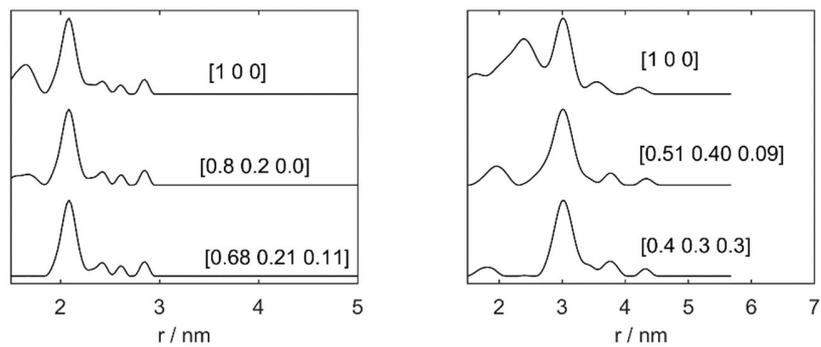

**Figure S3**. Distances Gd-ruler-2.1 (left) and Gd-ruler 3.0 (right) determined for aqueous protonated solutions with different overtone coefficients. From top to bottom: no overtones, overtone coefficients determined in a deuterated aqueous solution, overtone coefficients determined in a protonated aqueous solution.

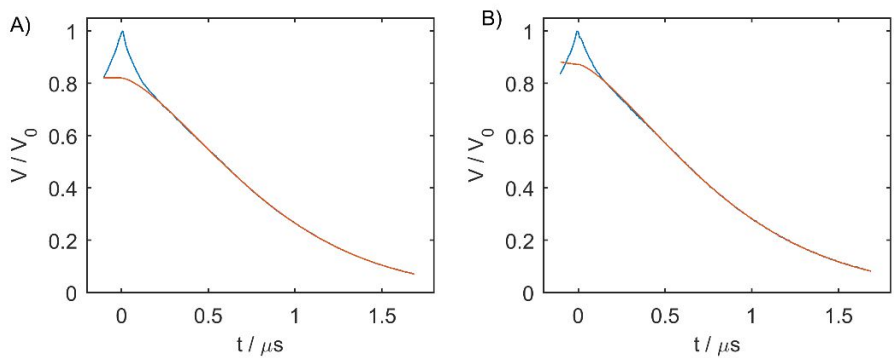

**Figure S4**. RIDME time traces (blue) and the background function (red) for Gd-ruler-3.0 in cell extract (A) and in oocytes (B).

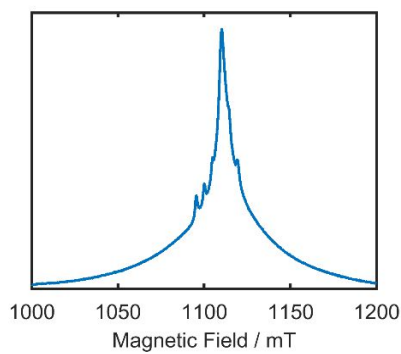

**Figure S5**. Echo-detected field sweep of Gd-ruler-3.0 in oocytes, recorded at 10K.

**References:**

1. Qi, M.; Hülsmann, M.; Godt, A. Spacers for Geometrically Well-Defined Water-Soluble Molecular Rulers and Their Application. *J. Org. Chem.* **2016**, *81(6)*, 2549-2571.

2. Clayton, J. A.; Qi, M.; Godt, A.; Goldfarb, D.; Han, S.; Sherwin, M. S. $Gd^{3+}$-$Gd^{3+}$ Distances Exceeding 3 nm Determined by Very High Frequency Continuous Wave Electron Paramagnetic Resonance. *Phys. Chem. Chem. Phys.* **2017**, *19*, 5127-5136.

3. Qi, M.; Groß, A.; Jeschke, G.; Godt, A.; Drescher, M. Gd(III)-PyMTA Label Is Suitable for In-Cell EPR. *J. Am. Chem. Soc.* **2014**, *136*, 15366-15378.

4. Milikisyants, S.; Scarpelli, F.; Finiguerra, M. G.; Ubbink, M.; Huber, M. A Pulsed EPR Method to Determine Distances between Paramagnetic Centers with Strong Spectral Anisotropy and Radicals: The Dead-Time Free RIDME Sequence. *J. Magn. Reson.* **2009**, *201*, 48-56.

5. Keller, K.; Mertens, V.; Qi, M.; Nalepa, A. I.; Godt, A.; Savitsky, A.; Jeschke, G.; Yulikov, M. Computing Distance Distributions from Dipolar Evolution Data with Overtones: RIDME Spectroscopy with Gd(III)-based Spin Labels. *Phys. Chem. Chem. Phys.* **2017**, *19*, 17856-17876.

6. Jeschke, G.; Chechik, V.; Ionita, P.; Godt, A.; Zimmermann, H.; Banham, J.; Timmel, C. R.; Hilger, D.; Jung, H. DeerAnalysis2006 – A comprehensive Software Package for Analyzing Pulsed ELDOR Data. *Appl. Magn. Reson.* **2006**, *30 (3-4),* 473-498.